\documentclass[conference]{IEEEtran}
\IEEEoverridecommandlockouts
\usepackage{url}
\usepackage{cite}
\usepackage{amsmath,amssymb,amsfonts}
\usepackage{algorithmic}
\usepackage{graphicx}
\usepackage{textcomp}
\usepackage{xcolor}
\usepackage{NotationMacros}
\usepackage{booktabs}

\def\BibTeX{{\rm B\kern-.05em{\sc i\kern-.025em b}\kern-.08em
    T\kern-.1667em\lower.7ex\hbox{E}\kern-.125emX}}
\begin{document}

\title{Sensitivity of Room Impulse Responses in Changing Acoustic Environment}

\author{\IEEEauthorblockN{Karolina Prawda}
\IEEEauthorblockA{\textit{AudioLab, School of Physics, Engineering and Technology, } \\
\textit{University of York, York, UK} \\
karolina.prawda@york.ac.uk}}

\maketitle

\begin{abstract}
Changes in room acoustics, such as modifications to surface absorption or the insertion of a scattering object, significantly impact measured room impulse responses (RIRs). These changes can affect the performance of systems used in echo cancellation and active acoustics and support tasks such as navigation and object tracking. Recognizing and quantifying such changes is, therefore, critical for advancing technologies based on room acoustics. This study introduces a method for analyzing acoustic environment changes by evaluating the similarity of consecutively recorded RIRs. Short-time coherence is employed to characterize modifications, including changes in wall absorption or the presence of a moving person in the room. A sensitivity rating is further used to quantify the magnitude of these changes. The results clearly differentiate between types of modifications---atmospheric variation, changes in absorption, and human presence. The methods described provide a novel approach to analyzing and interpreting room acoustics, emphasizing RIR similarity and extracting information from temporal and spectral signal properties.
\end{abstract}


\section{Introduction}\label{sec:intro}
The acoustics of a room are typically characterized by parameters such as reverberation time (RT), clarity, and definition, which are derived from room impulse responses (RIRs) \cite{vorlnder1998objective, patynen2013}. However, these objective measures do not fully capture the variability of RIRs over time and space \cite{PELORSON1992_variability}. Consequently, multiple measurements are often required to understand spatio-temporal changes in acoustic conditions. This study proposes an alternative approach to analyzing RIRs in time-varying environments.

Changes in room geometry and acoustic properties introduce differences in RIRs \cite{Guski_2015_scattering, becker_1984_echos, connor1973experimental, Green2016ASSESSINGTS}, which depend on the specific nature of the modification. Understanding such changes in room transfer functions is essential for designing active acoustics and echo cancellation systems, as these systems must be precisely calibrated to avoid feedback loop instability \cite{connor1973experimental, Green2016ASSESSINGTS, antweiler_1995_simulation, Elko:2003_Thermal}. Analyzing RIRs in response to room geometry changes can aid in identifying and tracking modifications, such as the presence and movement of people \cite{wang2023soundcam, MARMAROLI2023109161}. Additionally, acoustic signals can enhance the performance of tracking and navigation systems for mobile robotics, particularly in environments with transparent surfaces \cite{singh2019multi} or occluding objects \cite{Lindell_2019_CVPR}. In all these applications, it is crucial to distinguish stochastic variations in the measurement environment \cite{prawda2024_coherence} from changes in its properties \cite{prislan2014} to ensure robust system calibration and reliable result analysis.

To address this, Prislan \emph{et al.} introduced a cross-correlation-based sensitivity measure to quantify differences in RIRs recorded in the same room with altered sound source positions \cite{prislan2014}. Their experiments revealed that high frequencies are more sensitive to changes in measurement conditions. Similarly, Satoh used cross-correlation to attribute changes in repeated RIRs to factors such as air temperature drift or air movement \cite{satoh2007}. More recently, Prawda \emph{et al.} extended this concept using frequency-dependent short-time coherence \cite{prawda2024_coherence}. Her work demonstrated that RIR variability, caused by the inherent instability of the propagation medium, increases with frequency and RIR duration \cite{prawda2024_coherence}.

This study introduces a method to analyze differences between repeated RIRs when substantial changes occur in the measurement environment. Short-time coherence is employed to characterize temporal and frequency-domain changes, while a sensitivity rating provides a single-parameter descriptor of these modifications.

The paper is organized as follows: Section~II describes the RIR analysis methodology. Section~III presents the results of applying the proposed approach to measurements with varying surface absorption, while the effect of a moving person is discussed in Section~IV. Section~V concludes the study.







\section{Methodology} \label{sec:method}

Differences in consecutively measured RIRs in the same space can stem from several factors: time variance, manifested, for example, through changes in atmospheric conditions and medium movements \cite{prawda2024_coherence, TRONCHIN2021107933, OTHMANI2023119953}, and modifications in the acoustics of the space, such as when objects are added or moved within the room \cite{cucharero2019_placement, gotz2021_motus, prawda2022calibrating, becker_1984_echos}. Here, the focus is on the latter. The discussion revolves around the coherence between repeated RIR measurements and the related sensitivity rating.

\subsection{Short-time coherence}
Two repeated acoustic measurements are modeled as follows:
\begin{equation}
\begin{aligned}
    \x\tf = \sirx\tf + \nx \tf, \\
    \y\tf = \siry\tf + \ny \tf,
    \label{eq:xi}
\end{aligned}
\end{equation}
where $\sirx$ and $\siry$ are RIRs and $\nx$ and $\ny$ are stationry background noise terms at time $t$ and frequency $f$. It is assumed that $\sirx$ and $\siry$ have the same distribution and analogously $\nx$ and $\ny$. The coherence between $\x$ and $\y$ is
\begin{equation}
    \PCC\tf = \frac{\abs*{\EXb{(\sirx + \nx)\compconj{(\siry + \ny)}}}^2}{\EXb{\abs{\sirx + \nx}^2} \EXb{\abs{\siry + \ny}^2}},
    \label{eq:PCC}
\end{equation}
where $\EX$ is the expected value across time and $\compconj{()}$ denotes the complex conjugate. For discrete signals, $\EX$ is approximated by a short-time average. Here, the time-frequency dependency is omitted where appropriate for the sake of conciseness.

$\PCC\tf$ can be decomposed into two terms: the loss of coherence due to changes in the measurement environment, $\PCCir\tf$, and the expected loss of coherence resulting from sound energy decay and increasing contribution of uncorrelated background noise \cite{prislan2014, prawda2022ro2, prawda2024_coherence}, $\PCCexp\tf$:
\begin{equation}
\begin{aligned}
   &\PCC\tf = \PCCir\tf \PCCexp\tf,\\
\end{aligned}
\label{eq:snrxtv}	
\end{equation}
where
\begin{subequations}

\begin{equation}
\begin{aligned}
 & \PCCir\tf = \Biggl(\frac{\abs*{\EXb{\sirx \compconj{\siry}}}}{\EXb{\abs{\sirx}^2}}\Biggr)^2, \\   
\end{aligned}
\label{eq:tv}	
\end{equation}
\begin{equation}
\begin{aligned}
&  \PCCexp\tf = \Biggl(\frac{\Esir}{\Esir + \En} \Biggr)^2~.\\   
\label{eq:snr}	
\end{aligned}
\end{equation}
\end{subequations}

Previous research \cite{prawda2024_coherence} shows that in the case of stochastic changes that affect all the RIR reflections (temperature fluctuations, turbulence in the propagation medium), the effect of atmospheric variation can be assessed independently. However, when modified acoustics result in different reflection amplitudes (due to absorption changes), changes in the nature of the reflections (scattering changes), or the absence of some reflections (occlusion) in a RIR, disentangling the effects in \eqref{eq:tv} becomes significantly more complex. Tracking modified reflections is particularly challenging when the RIRs become increasingly diffuse. To address this, the present study focuses on a holistic assessment of the impact of room acoustics on coherence and aims to provide empirical insights into this issue.
 
\subsection{Sensitivity rating}

To quantify the amount of change that has occurred within the acoustic environment under test, we use the sensitivity rating $\Gamma$ as a single-parameter descriptor \cite{prislan2014}. Unlike correlation, $\Gamma$ quantifies how different or uncorrelated two RIRs are. It assumes values between~0 and~1, where 0 indicates that the RIRs are not sensitive to variations in the measurement environment and thus identical, and 1 indicates that the changes greatly affect the measured RIRs, prompting the signals to become completely uncorrelated between the measurements.

In this study, the original definition of $\Gamma$ is used, however, the sensitivity function from \cite{prislan2014} is replaced with the short-time coherence
\begin{equation}
    \Gamma(\freq) = \frac{ \sum_t (1-\PCC\tf) \sqrt{\EXb{\abs{\sirx + \nx}^2} \EXb{\abs{\siry + \ny}^2}}}{\sum_t \sqrt{\EXb{\abs{\sirx + \nx}^2} \EXb{\abs{\siry + \ny}^2}}}~.
    \label{eq:sensitivity}
\end{equation}
Here, $1-\PCC\tf$ is non-negative, while the sensitivity function can be below zero, e.g., when the phases of reflections between a pair of RIRs are reversed. As this study focuses on the presence, amplitude and time-of-arrival of reflections, using coherence allows for easier analysis of frequency-dependent $\Gamma$ and offers robustness towards negative values.

When computing \eqref{eq:sensitivity}, attention must be paid to the summation limit and to analyzing only the useful RIR portions, i.e., those with sufficiently high SNR \cite{prislan2014}. Thus, similarly to \cite{prawda2024_coherence}, the sensitivity rating in this study is calculated for the parts of the RIRs that have an SNR of 30 dB or higher.



\begin{figure}
    \centering
    \includegraphics[clip, width=.95\linewidth]{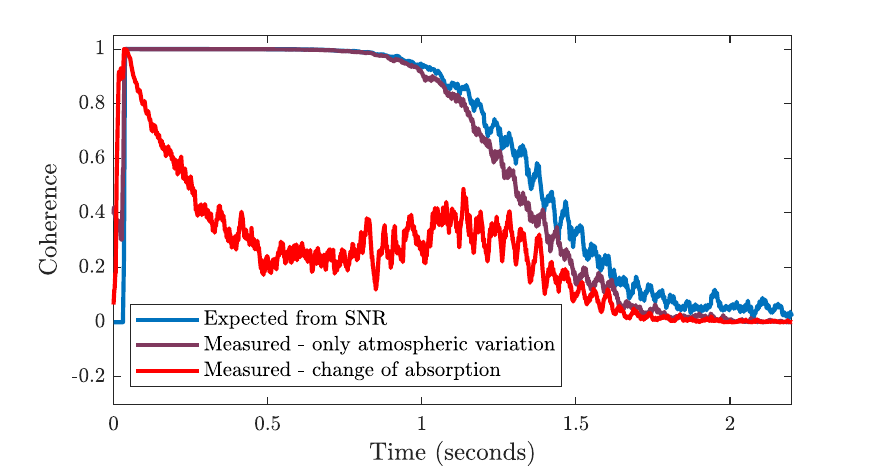}
    \caption{Short-time coherence between repeated RIRs from \emph{Arni} database. The coherence loss resulting from time-variant measurement environment is more similar to the expected coherence curve obtained through \eqref{eq:snr} than when the absorption distribution in the room has changed.}
    \label{fig:Arni_52_broad}
\end{figure}

\section{Change in surface absorption}\label{sec:absorption}

First, the effect of changing surface absorption was studied using the proposed methodology on a set of RIRs measured in the variable acoustic laboratory \emph{Arni} \cite{prawda_2022_Arni_Data}. \emph{Arni} features 55 variable acoustic panels, which can be configured in either a closed-reflective or open-absorptive state. Details about the room and the measurement procedures can be found in \cite{prawda2022calibrating}. In this study, the focus was placed on broadband coherence and sensitivity rating.

This experiment considered the combinations of RIRs with the same room equivalent absorption area $A~=~\sum_i~\alpha_i~S_i$, where $\alpha_i$ is the absorption coefficient and $S_i$ is the surface area of the $i$th wall. While the total absorption was constant across RIRs, the distribution varied, as the number of absorptive and reflective panels remained the same, but their placement in the room differed for each measurement.

The broadband short-time coherence of two reverberant combinations (\#5143 and \#5144 \cite{prawda_2022_Arni_Data}) in \emph{Arni}, each with 52 panels closed (RT = 1.2\,s). Figure~\ref{fig:Arni_52_broad} compares their coherence with that measured for RIRs subject to only atmospheric variation (the same panel configuration, repeated measurements) and the expected coherence loss calculated from \eqref{eq:snr}. Both the SNR-motivated and atmospheric variation-based coherence losses were gradual, accelerating only late in the RIR. In contrast, the coherence for the absorption change dropped rapidly to around 0.3, stabilizing until the sound energy decayed. This sharp distinction highlights how changes in room acoustics are clearly identifiable and easily distinguished from stochastic variations in RIRs.

The sensitivity rating $\Gamma$ for the example from Fig.~\ref{fig:Arni_52_broad} is 0.096 for the absorption distribution change and 0.0002 for atmospheric variation only.  This indicates that the RIRs are nearly 500 times more sensitive to changes in surface absorption than to the inherent instability of the measurement environment.

\begin{figure}
    \centering
    \includegraphics[trim=0.0cm 1cm 0.0cm 0cm,  width=.95\linewidth]{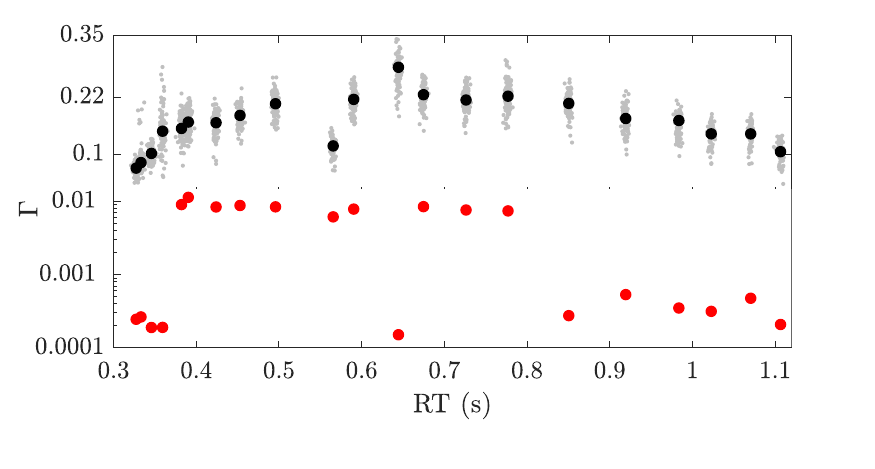}
    \caption{Sensitivity of RIRs from \emph{Arni} database. Grey dots show single values of $\Gamma$ for each RIR pair, whilst black dots represent median values for each analyzed condition of $A$. The red dots are the sensitivity values for when the change in RIRs are caused only by atmospheric variation. Note that the top half of the Y-axis is linear, while the bottom half is logarithmic for enhanced readability of the results.}
    \label{fig:Arni_sens}
\end{figure}

To further investigate the short-time coherence and sensitivity ratings for changes in the measurement environment within \emph{Arni}, several $A$ values were analyzed. For each open-closed panel configuration, all 100 measured RIRs were utilized, with the first RIR from each condition serving as the reference. The remaining 99 signals were correlated with the reference, and $\Gamma$ values were calculated for each coherence curve.

Figure~\ref{fig:Arni_sens} shows the sensitivity rating for the considered conditions, ordered according to their RT values. Additionally, a median sensitivity rating was computed for each condition.

The results reveal a rise in $\Gamma$ as RT increases, peaking around the midpoint of the total RT range, followed by a subsequent decline. This trend aligns with the extent of absorption distribution change in each condition. At low RTs, only a few panels transition from reflective to absorptive states between consecutive RIRs (e.g., three out of 55 panels on the left side of the X-axis in Fig.~\ref{fig:Arni_sens}), resulting in minimal absorption distribution variation. Similarly, at high RTs, where the majority of panels are reflective (e.g., 53 out of 55 panels on the right side of the X-axis), the room's absorption distribution shows limited variability. However, at intermediate RTs, where 20 to 30 out of 55 panels switch states between RIR measurements (middle of the X-axis), the absorption distribution changes significantly, yielding higher sensitivity ratings. Thus, $\Gamma$ not only identifies the source of change in room acoustics but also quantifies the extent of that change.

Figure~\ref{fig:Arni_sens} also compares $\Gamma$ values for absorption-induced changes and atmospheric-variation-induced changes between RIRs, where the same panel configuration was measured multiple times. The results corroborate those from Fig.~\ref{fig:Arni_52_broad}, showing that $\Gamma$ for atmospheric variation is orders of magnitude lower than for absorption changes, making these variations easily distinguishable. Moreover, $\Gamma$ resulting from atmospheric variation is independent of RT or panel configuration changes but is instead linked to the environmental instability during the measurement. For instance, measurements with RTs between 0.38\,s and 0.8\,s were captured in conditions with greater variability, such as stronger air movement, compared to other measurements.

\begin{figure}[t!]
    \centering
    \includegraphics[trim=0.0cm 0.10cm 0.0cm 0cm, clip, width=0.95\linewidth]{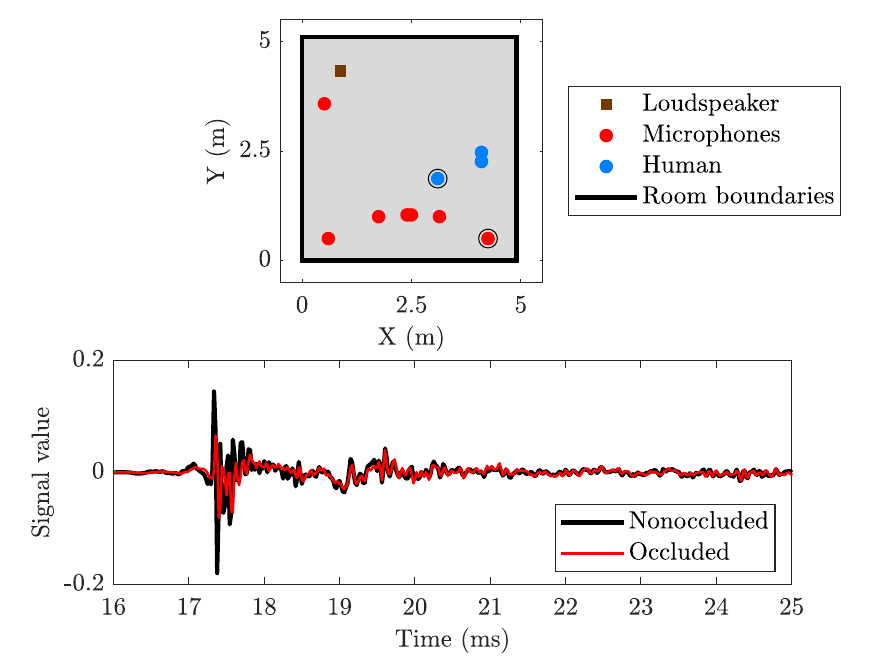}
    \caption{(Top) Layout of the analyzed room from the SoundCam database. The circled microphone position is occluded in one of the considered configurations, when the human is in the circled position. (Bottom) A close-up of first 10\,ms of two RIRs measured with the circled microphone: (black) direct path nonoccluded, (red) direct path occluded.  }
    \label{fig:Sc_rir}
\end{figure}

\section{Human presence in the room}\label{sec:human}
In addition to analyzing the impact of absorption distribution, this study examined the effect of a human presence on RIR sensitivity. A person introduces scattering and absorption into the measurement space \cite{conti2024_human} and may occlude certain microphones when positioned in the line of sight between the sound source and receiver. To investigate this, we utilized the SoundCam database \cite{wang2023soundcam}, which contains RIR measurements from various rooms, both empty and with a human present who changes location between recordings. 

Here, we used the RIRs captured in the \emph{Treated Room} scenario. The top pane of Fig.~\ref{fig:Sc_rir} depicts the measurement space with the sound source, receiver, and human's positions in three consecutive measurements. In one of those, a microphone is occluded, as marked by the circle around the microphone and the occluder positions. The bottom pane of Fig.~\ref{fig:Sc_rir} illustrates an example of two RIRs from the \emph{Treated Room}. The direct path between the sound source and the microphone is occluded in one of the RIRs, causing considerable differences compared to the nonoccluded RIR, especially in the direct sound and the first few reflections.

To account for the human body’s frequency-dependent effects on sound waves, $\PCC$ and $\Gamma$ were analyzed across different frequency bands. A constant bandwidth of 1\,kHz was chosen, centered ±500\,Hz around frequencies ranging from 1 to 19\,kHz \cite{prawda2024_coherence}, to ensure adequate resolution at high frequencies.

\begin{figure}
    \centering
    \includegraphics[trim=0cm 0.35cm 0cm 0cm, clip, width=0.95\linewidth]{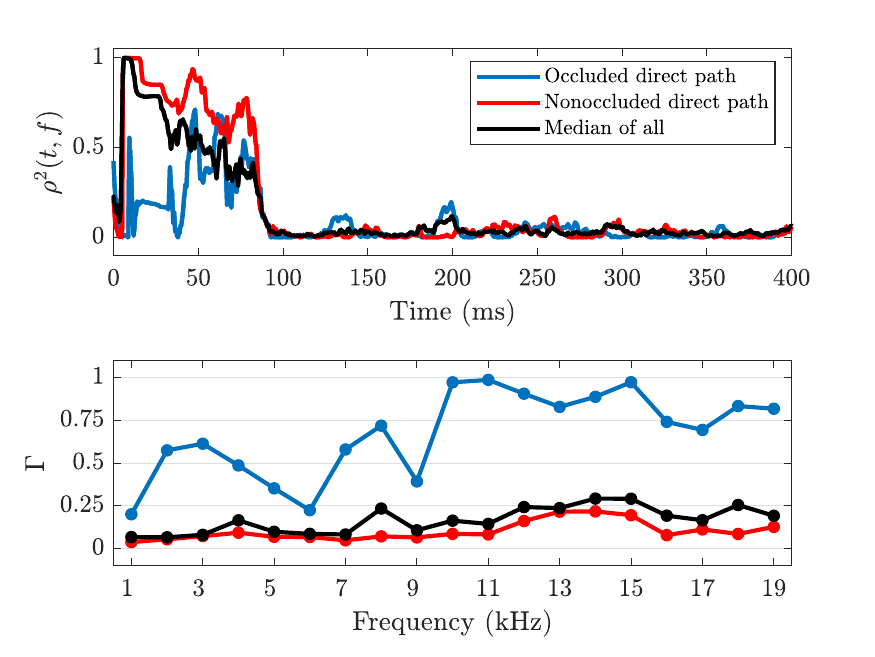}
    \caption{(Top) Short-time coherence curves for occluded (blue) and nonoccluded (red) microphones compared to the median of all receivers for the 19-kHz band. (Bottom) Corresponding $\Gamma$ for frequency bands between 1--19\,kHz. }
    \label{fig:Sc_med}
\end{figure}

The top pane of Fig.~\ref{fig:Sc_med} illustrates the short-time coherence values for the 19-kHz band. When the direct path is occluded in one RIR of a pair, the coherence curve drops to near-zero values at the beginning but increases over time. This pattern allows for straightforward identification of the person’s position, as the timing of low-coherence values corresponds to the reflections' time-of-arrival and the room geometry.

For the assessment of the total change in room acoustics, however, a near-zero coherence may introduce out-of-distribution values that destabilize the analysis and the resulting sensitivity rating. Comparing the coherence curves for occluded and nonoccluded direct paths in the top pane of Fig.~\ref{fig:Sc_med}, the estimation of the effect of human position during this particular measurement session is overestimated by looking just at the results from the occluded microphone. Therefore, we propose to use the median of the short-time coherence of all the microphones that simultaneously recorded the same scene from different positions. The median is a stable estimate that is hardly affected by outliers \cite{donoho1983notion, LEYS2013764, rousseeuw1993alternatives, prawda2022ro2, prawda2024_mosaic}, therefore it allows to more accurately assess the room-wide effect in place of focusing on a specific sound path. In the top pane of Fig.~\ref{fig:Sc_med}, the median $\PCC$ positions itself between the coherence curves for the occluded and nonoccluded RIRs.

A similar pattern is visible in the sensitivity ratings, depicted in the bottom pane of Fig.~\ref{fig:Sc_med}. The $\Gamma$  values for the RIRs captured with the occluded microphone are much higher than those for the nonoccluded microphone, reaching close to unity in the 10, 11, and 15-kHz bands. Similar to the short-time coherence, the sensitivity rating helps to estimate the location of the occluder in the room, albeit less accurately due to the lack of a direct connection between $\Gamma$ and room geometry. Again, using the median sensitivity rating allows for an estimation of the room-wide effect.

\begin{figure}
    \centering
    \includegraphics[trim=0cm 0.40cm 0cm 0cm, clip, width=0.95\linewidth]{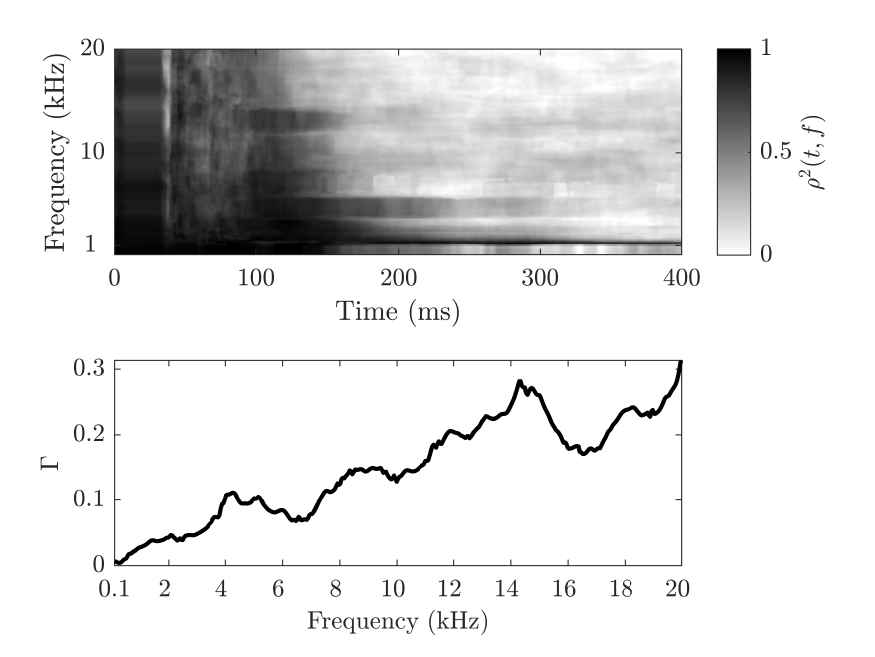}
    \caption{(Top) Time-frequency short-time coherence for two RIRs from SoundCam dataset \cite{wang2023soundcam}. (Bottom) Corresponding frequency-dependent sensitivity $\Gamma$.}
    \label{fig:SoundCam_spec}
\end{figure}

To obtain more detailed information about the time-frequency characteristics of the short-time coherence and sensitivity rating, we calculate~\eqref{eq:PCC} and~\eqref{eq:sensitivity} on short-time Fourier transform of a pair of RIRs with nonoccluded direct path. The results are displayed in Fig.~\ref{fig:SoundCam_spec}.

The top pane of Fig.~\ref{fig:SoundCam_spec} shows the temporal evolution of the short-time coherence for frequencies linearly spaced between 100\,Hz and 20\,kHz. The initial high values of $\PCC$ are followed by a sudden drop around 38\,ms, which is consistent with the coherence plots in the top pane of Fig.~\ref{fig:Sc_med}, and might result from the difference in the reflection pattern due to the changed position of a human between the two measurements. The drop widens with increasing frequency, suggesting a stronger effect of the occluder on shorter wavelengths.

The sensitivity rating in the bottom pane of Fig.~\ref{fig:SoundCam_spec} shows a nearly linear increase in $\Gamma$ over frequency and is similar to the values for nonoccluded microphone in Fig.~\ref{fig:Sc_med}, although differences associated with the bandwidth appear \cite{prislan2014}. The level of detail in the high frequencies over 10 kHz justifies the use of linear frequency spacing and constant bandwidth over the more common logarithmic approach.

\section{Conclusion}
\label{sec:conclusion}

This paper presents a method to analyze the differences between repeated RIR measurements, focusing on the changes in absorption distribution in the room and the presence of scattering and absorbing object, in this case, a human, which changes location between consecutive recordings. The results demonstrate that using short-time coherence and sensitivity rating effectively indicate changes in the measured conditions and quantify the extent of these changes. The proposed methodology is a valuable analysis tool in application scenarios such as acoustic echo cancellation, active acoustic systems, and other areas that require detailed knowledge of the acoustics of the measurement environment and its stability.

\clearpage 
\bibliographystyle{IEEEbib}
\bibliography{JASA_variable}

\end{document}